\documentclass[twocolumn,floatfix, showpacs, prc]{revtex4-1}

\usepackage[final]{epsfig}
\usepackage{amsmath}
\begin{document}
 
\title{Constraints from $v_2$ fluctuations for the initial state geometry of
heavy-ion collisions}
 
\author{Thorsten Renk}
\author{Harri Niemi}
\affiliation{Department of Physics, P.O. Box 35, FI-40014 University of
Jyv\"askyl\"a, Finland}
\affiliation{Helsinki Institute of Physics, P.O. Box 64, FI-00014 University of
Helsinki, Finland}

 
\begin{abstract}
The ability to accurately compute the series of coefficients $v_n$ characterizing the momentum space anisotropies of particle production in ultrarelativistic heavy ion collisions as a function of centrality is widely regarded as a triumph of fluid dynamics as description of the bulk matter evolution. A key ingredient to fluid dynamical modeling is however the initial spatial distribution of matter as created by a yet not completely understood equilibration process. A measurement directly sensitive to this initial state geometry is therefore of high value for constraining models of pre-equilibrium dynamics. Recently, it has been shown that such a measurement is indeed possible in terms of the event by event probability distribution of the normalized $v_n$ distribution as a function of centrality, which is to high accuracy independent on the details of the subsequent fluid dynamical evolution and hence directly reflects the primary distribution of spatial eccentricities. We present a study of this observable using a variety of Glauber-based models and argue that the experimental data place very tight constraints on the initial distribution of matter and rule out all simple Glauber-based models.  
\end{abstract}
 
\maketitle

\section{Introduction}

It is now commonly agreed that ultrarelativistic heavy-ion (A-A) collisions 
create a transient state of collective QCD matter. In modeling the dynamics 
of this droplet, the essential input to the models is an initial 
distribution of matter density as created in a yet not completely understood 
equilibration process~\cite{Strickland:2013uga}. One of the clearest signal of collective (or fluid dynamical) behavior
of such a system is the appearance of non-trivial patterns in the azimuthal
distribution of final state hadron spectra~\cite{Heinz:2013th}. Such patterns are created by the fluid dynamical
response to pressure gradients which in turn are given by the geometric shape of the initial state~\cite{Ollitrault_elliptic, Kolb:2003dz}. 
The precise details of this response then depend strongly on the transport properties of the matter, e.g. 
shear viscosity~\cite{Romatschke:2007mq,Luzum:2008cw,Schenke:2010rr,
Gale:2012rq,Song:2010mg,Song:2011qa,Shen:2010uy,Bozek:2009dw,Bozek:2012qs, 
Niemi:2011ix, Niemi:2012ry}.

The azimuthal asymmetries of the measured hadron momentum spectra are usually characterized by the set of
Fourier coefficients $v_n$, and similarly the azimuthal distribution of matter in position space
can be characterized by its eccentricity coefficients $\varepsilon_n$. The
determination of the viscosity of the strongly interacting matter is largely
based on measured $v_n$'s. However, $v_n$'s do not only depend on the fluid
dynamical response to $\varepsilon_n$'s , but also on the initial values of
$\varepsilon_n$'s. Therefore, it is essential that the right initial condition is used:
determining both the transport properties and the initial geometry
simultaneously from the available data is a very complicated task~\cite{Heinz:2009cv, Retinskaya:2013gca}. 
Thus, finding an observable that
is sensitive to the initial geometry, but independent of the fluid dynamical
response would simplify the task considerably. Another phenomena where the
detailed knowledge of the initial geometry becomes important is jet quenching,
where the observed azimuthal jet suppression patterns depend strongly on the assumed
initial state~\cite{Jetv2-1,Jetv2-2}.

In realistic modeling, the initial state geometry fluctuates from one collision
to the another even for a fixed impact parameter~\cite{Andrade:2006yh, Alver:2010gr}. 
The event-by-event fluctuations of $\varepsilon_n$'s then translate into the event-by-event
fluctuations of $v_n$'s. Fluid dynamical calculations have established that the
relation $\left<v_n\right> = C_n\varepsilon_n$, where the angular
brackets $\left<\right>$ denote the average over many collisions with the same
eccentricity, holds very well~\cite{Qin:2010pf, Qiu:2011iv, Gardim:2011xv, Niemi:2012aj}. 
For the second harmonics $v_2$ and $\varepsilon_2$ it has been found that the 
correlation is even stronger and a relation $v_2 = C_2 \varepsilon_2$ holds 
accurately also in individual nuclear collisions \cite{Niemi:2012aj},
not only on average. This means that in a given centrality class $\varepsilon_2$ is the
only characteristics of the initial condition that determines $v_2$, while the
proportionality coefficient $C_2$ depends on the details of the fluid dynamical
evolution in a complicated way~\cite{Teaney:2012ke}. The simple relation means that in relative
fluctuations $\delta v_2 = (v_2 - \left<v_2\right>)/\left<v_2\right>$ the
proportionality coefficient cancels. Therefore, the probability distribution
$P(\delta \varepsilon_n)$ is the same as the probability distribution $P(\delta
v_n)$. In other words $P(\delta v_n)$ is determined by the properties of initial
state alone and is independent of the fluid dynamical evolution. Thus, by
measuring $P(\delta v_n)$ one gets an immediate access to the fluctuations 
in the initial geometry~\cite{Niemi:2012aj}.

Recently, the event-by-event distributions of $v_n$ have been measured by the
ATLAS~\cite{ATLAS} and ALICE~\cite{ALICE} Collaborations. Making use of 
the result $P(\delta \varepsilon_n) = P(\delta v_n)$ we use different variants 
of the Monte-Carlo Glauber (MCG) model
to calculate $P(\delta \varepsilon_n)$, and compare with ATLAS
data. Furthermore, we study the sensitivity of the distributions to several 
assumptions underlying the MCG model and its extensions. 

\section{The model}

We compute the normalized fluctuations of $v_n$ by evaluating the spatial
eccentricity $\varepsilon_n$ of a set of randomly generated initial states for 
a given centrality class. 

We start by distributing the potentially interacting objects in the initial
states of the colliding nuclei. In the default scenario, these are the
nucleons, but in an alternative constituent quark scattering (CQS) scenario, 
we assume that the substructure of nucleons in terms of constituent quarks 
is the relevant level of description.

In the default case, we use a Woods-Saxon parametrization of the measured
nuclear charge density \cite{ChargeDist} to distribute nucleons randomly 
in a 3-dim volume. For Pb-nuclei as appropriate for the LHC, our distribution 
is given by
\begin{equation}
\label{E-WS}
\rho_N(r) = \frac{\rho_0}{1 + \exp ((r-c)/z)}
\end{equation} 
with $c = 6.61$ fm and the skin thickness $z = 0.51$ fm. We checked that
a slight changes in these parameters do not affect our result significantly.
We do not correct for the nucleon hard core, i.e.\ we permit configurations 
in which individual nucleons overlap in 3d space. After generating a 3d 
ensemble of nucleons, we project their position into transverse $(x,y)$ space. 
In the CQS scenario, we distribute three constituent quarks inside a Gaussian 
radius of $0.6$ fm around the nominal position of each nucleon, then project 
constituent quark positions into $(x,y)$ space.

In order to test alternative scalings, we also explore a Hard Sphere scenario (HS)
in which we set the skin thickness parameter $z=0$ and a Sheet (S) scenario in
which we mimick a strongly saturated picture in which we distribute nucleons a
priori into a 2d circular surface bounded by the nuclear radius parameter $c$
(in such a picture, the center of the nucleus is as dense as the periphery).
Both HS and S can be combined with the CQS scenario.

Once the transverse position of the colliding objects have been specified for
two nuclei, we displace the two distributions by a randomly sampled impact
parameter. Collisions are evaluated according to a transverse distance criterion
$d^2 < \sigma_{NN}/\pi$. In the case of nucleon-nucleon collisions we take
$\sigma_{NN} = 64$ mb, in the case of interacting constituent quarks we use
$1/9$ of this value to get back to the same cross section in the case of p-p
collisions.

There are four common ways in which EbyE hydrodynamics is commonly initialized.
Matter can be distributed either according to collision participants (wounded
nucleon, WN) or according to binary collisions (BC) and the matter distribution
can be specified in terms of entropy $s$ or energy density $e$, leading
to sWN, eWN, sBC and eBC scenarios. We note that none of these alone is able
to give a correct centrality dependence of the multiplicity.

For each event, we associate a binary collision and a wounded nucleon density
according to 
\begin{equation}
\rho(\mathbf{x})_{\rm bin/wn} = \sum_{i=1}^{N_{bin/wn}}\frac{1}{2\pi\sigma^2} \exp \left(-\frac{\mathbf{x}_i^2}{2\sigma^2} \right) 
\end{equation}
where $\mathbf{x}_i$ is the binary collision point or the position of the wounded nucleon,
and $\sigma$ is a free parameter. We will then consider three different 
possibilities to initialize the initial entropy density,
\begin{eqnarray}
\label{eq:bin}
s(\mathbf{x}) &=& N \rho_{\rm bc}(\mathbf{x})^\alpha, \\
\label{eq:wn}
s(\mathbf{x}) &=& N \rho_{\rm wn}(\mathbf{x})^\beta, \\
\label{eq:mix}
s(\mathbf{x}) &=& N \left[(f\rho_{\rm wn}(\mathbf{x}) + (1-f)\rho_{\rm bc}(\mathbf{x})\right],
\end{eqnarray}
where the parameters $\alpha$, $\beta$ and $f$ are fixed to reproduce the centrality
dependence of the multiplicity, by assuming that the final multiplicity is 
proportional to the initial entropy. We have tested that scaling $s \rightarrow s^{4/3}$,
corresponding to the approximate difference between $s$ and $e$ scaling,
does not change any of our results. 

In the default scenario, we set $N=1$, thus assuming that the multiplicity
created in each N-N collision is a constant. In real N-N collisions, the
multiplicity fluctuates and the relative distribution of multiplicity around the
mean value exhibits a near universal behaviour, the so-called KNO scaling
\cite{KNO}. In order to account for this, we also take into account a scenario
where $N$ is distributed according to the KNO distribution.

The value of $\sigma$ is characteristic for the interaction process, and
reflects the precise physics of matter production in secondary interactions.
General considerations suggest that it should be of the order of the nucleon
radius. We test in the following scenarios involving both constant values
$\sigma = 0.6$ fm, $\sigma = 1.0$ fm and a Gaussian distribution of width
$\Delta \sigma = 0.3$ fm centered around $\sigma = 0.6$ fm.

The events are divided into centrality classes according to the total
entropy, which is the closest to the centrality selection in the real experiments.
We further checked the sensitivity of the results to the centrality selection, by
considering also the selection according to impact parameter, number of collision
participants or the number of binary collisions.  It turns out that none of these
schemes to determine centrality changes our results substantially, i.e.\ the 
details of centrality determination do not matter for the question of $v_{n}$ fluctuations
as long as we do not consider ultra-central events.

Given $\rho(x,y)$, we compute the center of gravity of the distribution and
shift coordinates such that $(0,0)$ coincides with the center of gravity. Next
we determine the angular orientation of the $\varepsilon_n$ plane from
\begin{equation}
\Psi_n = \frac{1}{n} \arctan \frac{\int dx dy (x^2 + y^2) \sin (n\phi) 
\rho(x,y)}{\int dx dy (x^2 + y^2) \cos (n\phi) \rho(x,y)} + \pi/n
\end{equation}
and the eccentricity of the event as
\begin{equation}
\varepsilon_n = \frac{\int dx dy (x^2 + y^2) \cos [n (\phi - \Psi_n)]
\rho(x,y)}{\int dx dy (x^2 + y^2) \rho(x,y)}
\end{equation}

Averaging over a large number $O(20.000)$ of events, we determine the mean
eccentricity $\langle \varepsilon_n \rangle$ for each centrality class and express
the fluctuations in terms of the scaled eccentricity as
\begin{equation}
\delta \varepsilon_n = \frac{\varepsilon_n - \langle \varepsilon_n \rangle}{\langle
\varepsilon_n \rangle}
\end{equation}
where $\varepsilon_n$ is the eccentricity determined for a particular event.

\section{Results}

\subsection{Centrality dependence from the Glauber model}

\begin{figure*}[htb]
\epsfig{file=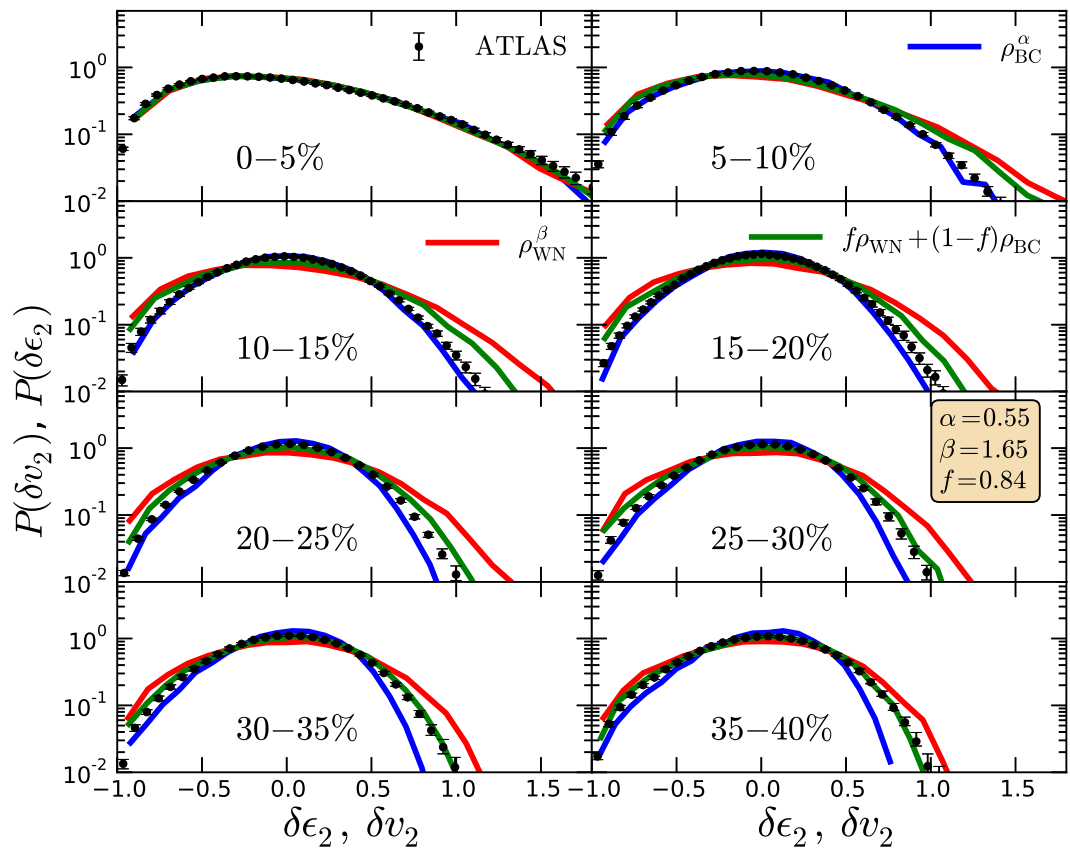, width=15cm}
\caption{Centrality dependence of $\delta v_2$ or $\delta
\varepsilon_2$ fluctuations in various scenarios to generate the initial state from the
initial nucleon distributions.}
\label{fig:e2_Glauber}
\end{figure*}

First, we test the centrality dependence of $P(\delta \varepsilon_2)$ of the different 
initial states given by Eqs.\ \eqref{eq:bin}--\eqref{eq:mix}. The distributions are
shown in Fig.\ \ref{fig:e2_Glauber} for several centrality classes and compared to 
the ATLAS data~\cite{ATLAS}.
The values of the free parameters $\alpha$, $\beta$ and $f$ are shown in the figure,
and we use $\sigma = 0.4$ fm and $N=1$. We can make the following observations:

\begin{itemize}
 \item In the most central collisions all the different models give the same
 distribution, and is in practice in perfect agreement with the ATLAS data.

 \item While it was observed in Ref.~\cite{Niemi:2012aj} that the distributions are same for 
 sWN and sBC initializations at RHIC, the same does not hold for the LHC energy
 due to the larger nucleon-nucleon cross-section. In general sWN initialization at the LHC 
 gives wider distributions than sBC initializations. This difference is even enhanced
 by the powers in Eqs.\ \eqref{eq:bin} and \eqref{eq:wn} required to reproduce the
 centrality dependence of the multiplicity.

 \item Although the binary collision based initialization, Eq.\ \eqref{eq:bin}, gives
 a good agreement with the data in the central collisions and the binary/participant 
 mixture, Eq.\ \eqref{eq:mix}, in the peripheral collisions, none of these simple models
 can fully account the centrality dependence of the distributions.
 
\end{itemize}

\subsection{Initial nuclear geometry}

Next, we aim at testing the assumptions for the initial nuclear geometry. In
particular we test four different scenarios across the whole centrality range:
1) a standard MC Glauber scenario based on nucleons distributed with a realistic
Woods-Saxon nuclear density (Glauber), 2) a standard Glauber scenario based on
scattering constituent quarks instead (CQS), 3) a Glauber scenario based on
nucleons sampled from a hard sphere distribution (HS) 4) a scenario mimicking
strong saturation effects in the initial density based on a 2d nucleon sheet 
distribution (S). In all these cases, $\sigma = 0.6$ fm and  $N=1$ is assumed.
Here, we use simple sBC model, with entropy density directly proportional
to the density of the binary collisions.

\begin{figure*}[htb]
\epsfig{file=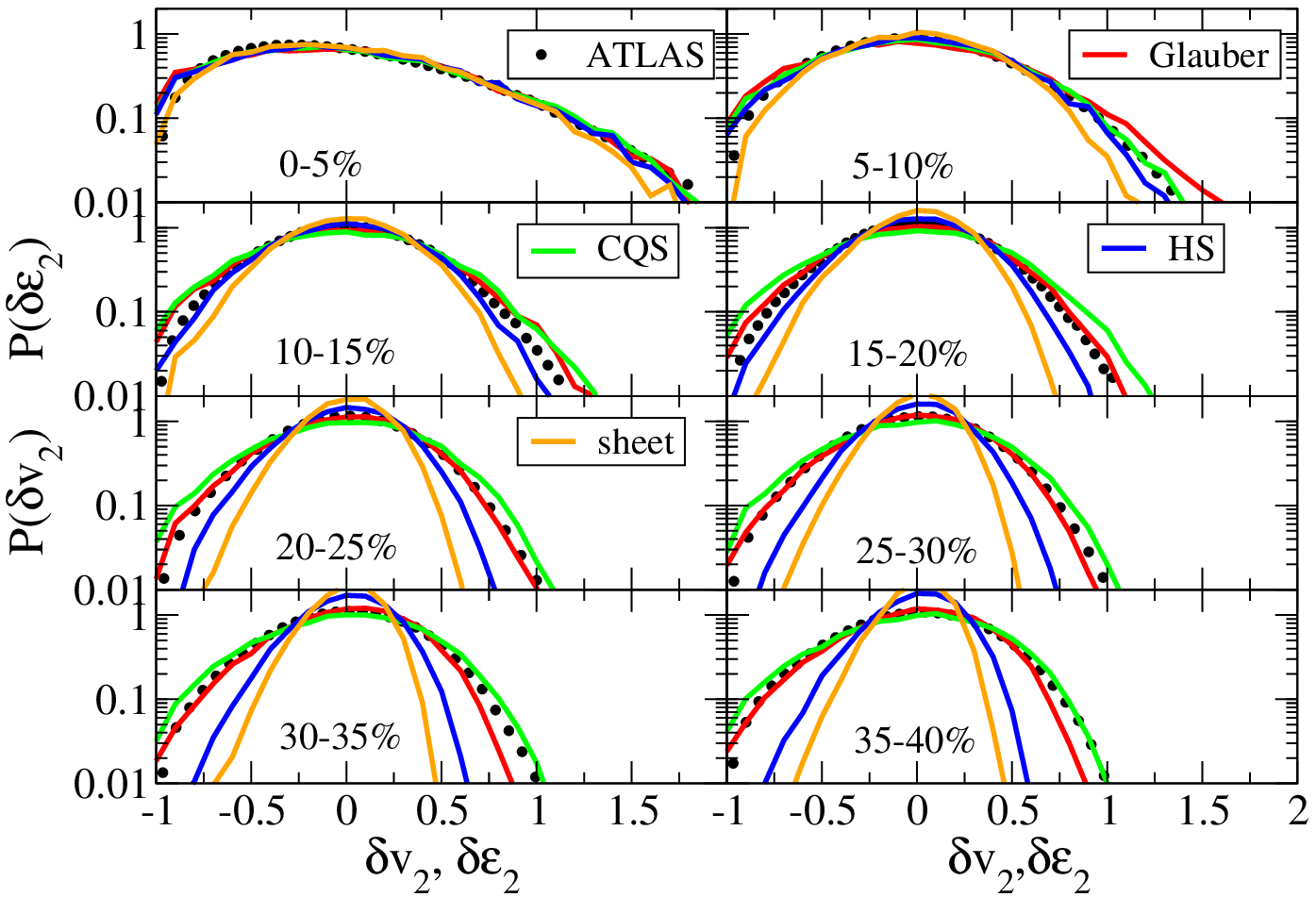, width=15cm}
\caption{\label{F-centrality}Centrality dependence of $\delta v_2$ or $\delta
\varepsilon_2$ fluctuations in various scenarios to generate the initial distributions
inside the nuclei (see text).}
\end{figure*}

The centrality dependence of the $v_2$ or $\varepsilon_2$ fluctuations from the
0-5\% most central to 35-40\% peripheral collisions is shown for these four
different scenarios and compared with ATLAS data in Fig.~\ref{F-centrality}.

Several observations are readily apparent:

\begin{itemize}
\item For central collisions, the scaled fluctuations in $v_2$ become universal,
i.e.\ show the same pattern independent of the underlying geometry. For less
central events, differences between the four different scenarios become readily
apparent.

\item As evidenced by the differences between Glauber and HS, the surface
diffuseness of the nucleus is a key parameter determining the width of the
distribution.

\item As indicated by the differences between HS and sheet, differences in
central density are also probed. This suggests a scenario in which the wide
fluctuations are driven by nucleons at the edge of one nucleus, passing (for
large impact parameters) through the central region of the other nucleus, 
i.e.\ what matters is both the probability to have a nucleon far from the center of
nucleus A and the effect of its passage through nucleus B. This would suggest
that any kind of saturation generically narrows the width of the distribution as
compared with an unsaturated scenario.

\item The more realistic scenarios Glauber and CQS are closer to the data, but
no scenario can account for the full centrality dependence. In particular,
Glauber becomes too narrow above 20\% centrality and CQS is too wide between 
10 and 25\% centrality.
\end{itemize}

We have similarly studied the centrality dependence of $v_3$ and $v_4$
fluctuations, however these appear to follow the same generic scaling as
observed for $v_2$ in central collisions and do not allow to distinguish
different models.

These results raise the question if a modified version of the Glauber scenario,
for instance size scale fluctuations or KNO multiplicity fluctuations could not
bring the model in agreement with the data. We explore these possibilities for
5-10\% centrality (where the Glauber scenario gives a wider distribution than the data) and
for 35-40\% centrality (where the width of the data is underestimated).

\subsection{Size scale and multiplicity fluctuations}

\begin{figure*}[htb]
\epsfig{file=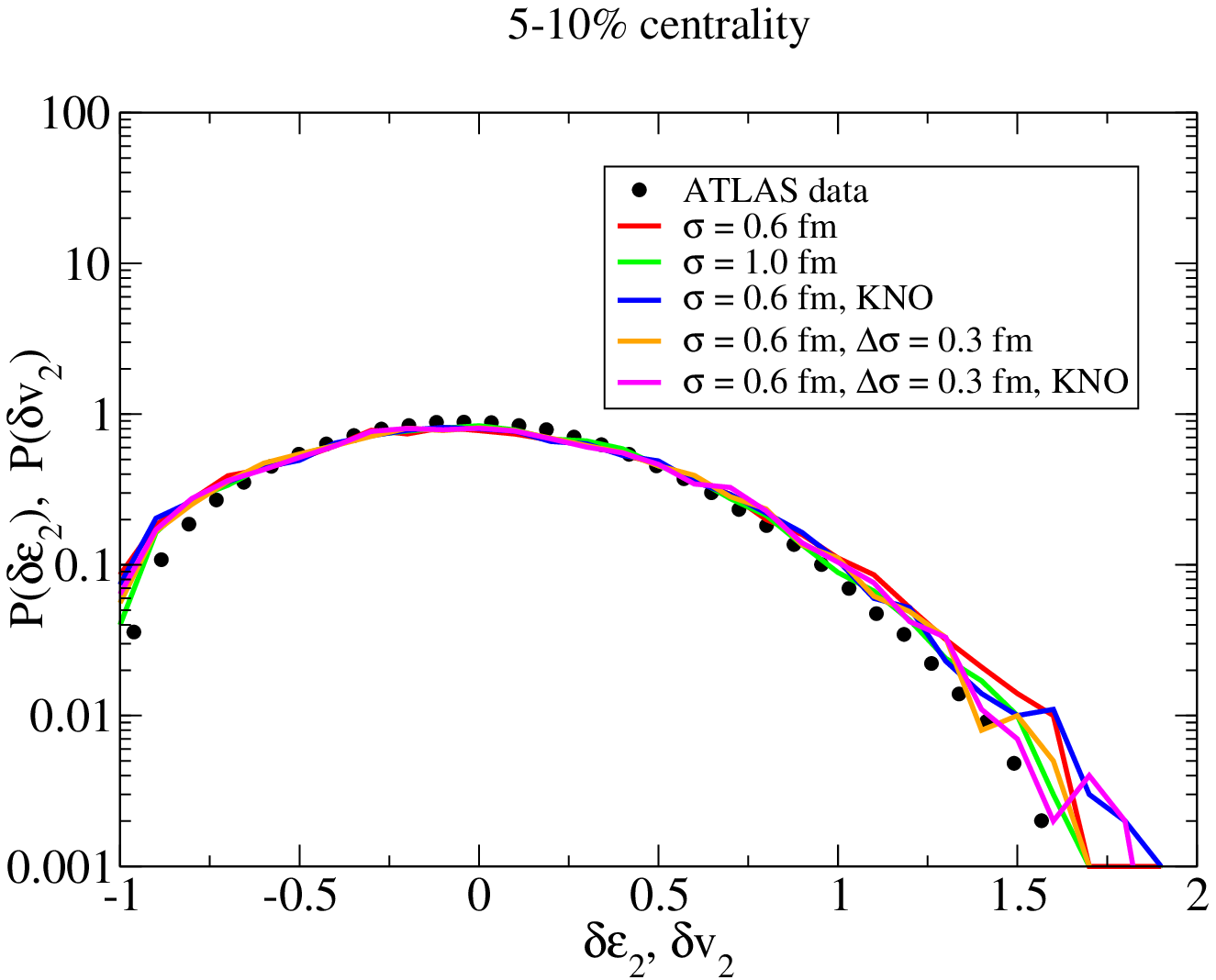, width=8.5cm}\epsfig{file=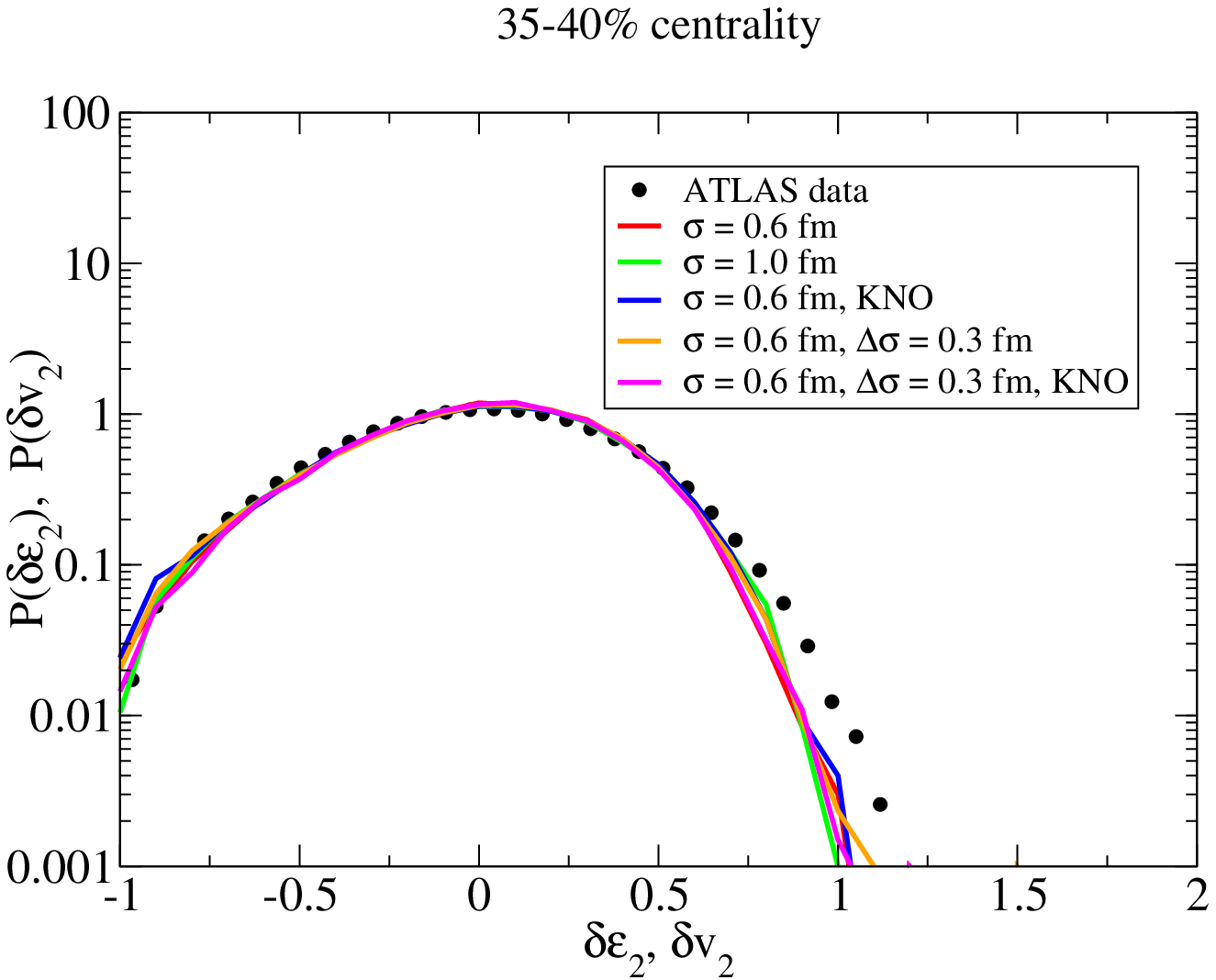,
width=8.5cm}
\caption{\label{F-Fluct} Dependence of $\delta v2$ fluctuations on multiplicity
or N-N collision geometry size scale fluctuations. } 
\end{figure*}

In Fig.~\ref{F-Fluct}, we again consider the sBC Glauber scenario and try
variations of the parameter $\sigma$ which represents the size of the matter
spot generated in an individual N-N collision in combination with possible
KNO-type multiplicity fluctuations. As the figure demonstrates, there is no
significant dependence on either of these factors, in particular no combination
of parameters is able to shrink the distribution at central collisions while at
the same time widen it at large centralities.

Similar results (not shown here) can be obtained for the CQS scenario.

\subsection{Surface thickness}

\begin{figure}
\epsfig{file=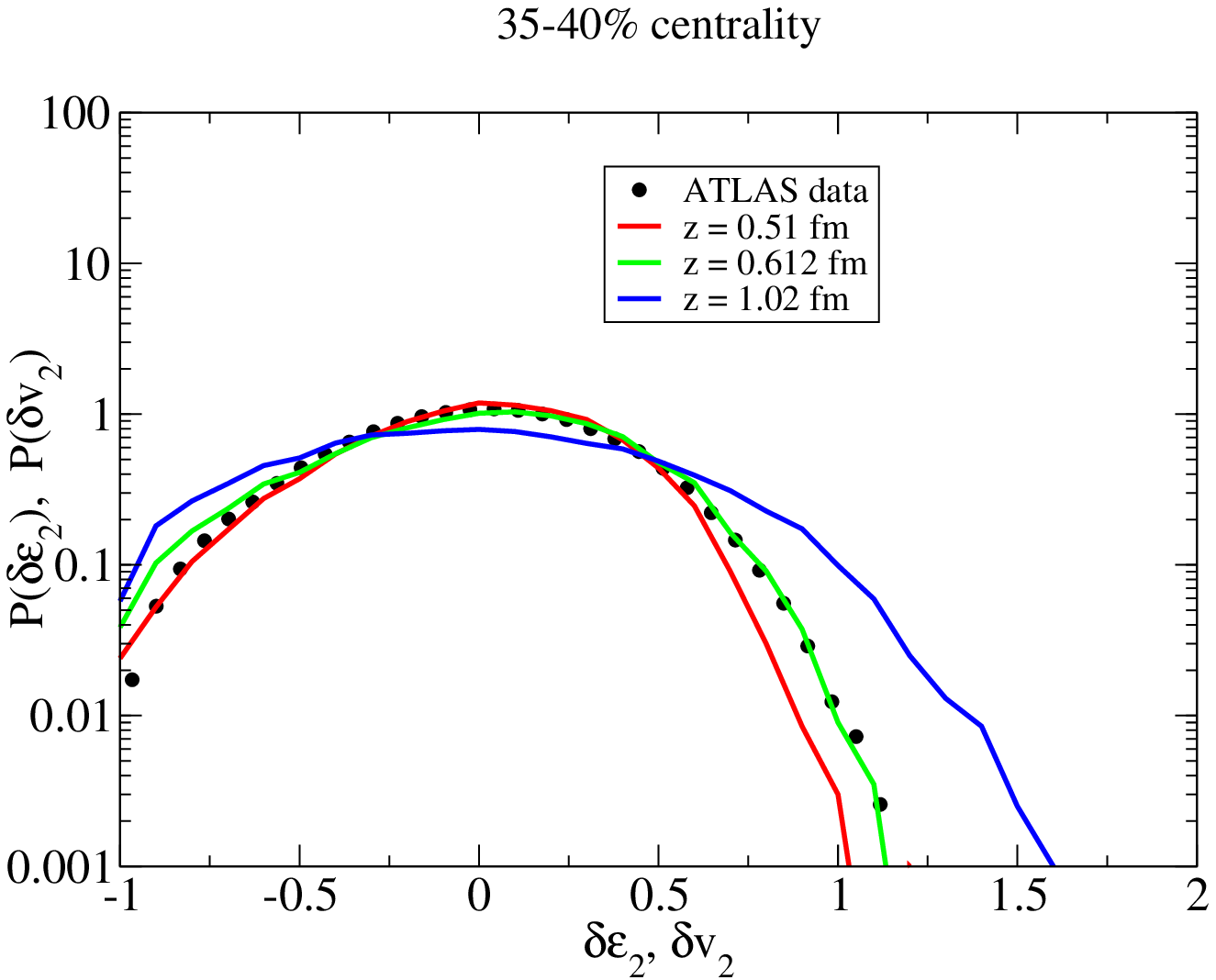, width=8.5cm}
\caption{\label{F-Skin} Dependence of $\delta v_2$ fluctuations on the assumed
surface diffuseness parameter in the Woods-Saxon model of the nuclear density
distribution.}
\end{figure}

In contrast, we demonstrate in Fig.~\ref{F-Skin} that there is a characteristic
dependence of the width of the $P(\delta v_2)$ on the surface diffuseness assumed
for the nuclear density distribution Eq.~(\ref{E-WS}) --- the distribution
widens with increased surface diffuseness and shrinks with decreased surface
diffuseness. Of all influences tested for a Glauber model based on colliding
nucleons, this is the only one clearly leading to an effect above the
statistical uncertainty.

However, even assuming yet unknown physics allows to make the surface
diffuseness a free parameter, there are two further obstacles:

\begin{itemize}
\item The surface diffuseness always correlates positively with the width of
$P(v_2)$. However, the centrality dependence of the mismatch between data and
model is non-trivial, i.e.\ in order to account for the data one would have to
assume that the surface diffuseness of a nucleus (which is a property of the
particular nucleus) depends on at what impact parameter that nucleus will later
collide, which is conceptually very problematic.

\item While $P(\delta v_2)$ is constrained to fulfill $\int d \delta v_2 P(\delta
v_2) = 1$ and $\int d \delta v_2 \delta v_2 P(\delta v_2) = 0$ by construction,
the shape is, given these constraints, free. Looking closely at
Fig.~\ref{F-Skin}, one may note that a different value of the surface
diffuseness reproduces the left and side and the right hand side of the
distribution, i.e.\ data and model do not match in shape. 

\end{itemize}

\section{Conclusions}

We calculated the centrality dependence of the eccentricity fluctuation spectra from 
several MC Glauber model based initial states. First, we found that the $v_2$ fluctuations 
are universal in the most central collisions, i.e.\ independent of the model details,
and well described by all the models. The same holds for the higher harmonics in 
all the centrality classes. However, none of the models tested here were able 
to reproduce the centrality dependence of $P(\delta v_2)$ observed by the ATLAS 
Collaboration. In particular, a simple mixture of binary collisions and wounded 
nucleons fails to reproduce the fluctuation spectra, except in the most peripheral 
centrality classes considered here. 

We also identified several parameters that do not affect the distribution, 
like KNO fluctuations and the size of the matter spots generated 
in the individual NN collisions. We further demonstrated that the distributions are 
sensitive to simple non-linear parametrizations given by Eqs.\ \eqref{eq:bin} 
and \eqref{eq:wn}, as well as by the changes in the initial distributions of 
the interacting objects.

These findings suggest that the geometrical fluctuations in the positions of the 
nucleons are not enough to explain the data, but some non-linear dynamics in the
creation of the matter and/or additional sources of fluctuations are necessary.
Both of these properties are realized in the QCD based initial state models.
For example, pQCD + saturation model in Ref.\ \cite{Eskola:2000xq, Paatelainen:2013eea} 
leads to a similar non-linear behavior of the entropy density as Eq. \eqref{eq:bin} and 
the sub-nucleon color fluctuations in Refs.\ \cite{CGCs} presumably 
lead to a similar effect on the distributions as the CQS model above.

Overall, reproducing the observed centrality dependence of the $v_2$ fluctuation 
distributions is a non-trivial task and gives very tight constraints for the
modeling of the initial state. All the simple models considered in this work 
can already be ruled out as valid representations of the initial state geometry.

\begin{acknowledgments}
 
We thank G.\ Denicol for useful discussions. 
This work is supported by the Academy researcher program of the Academy of Finland, 
Project No. 130472 and the Academy of Finland, Project No. 133005.
 
\end{acknowledgments}

\end{document}